# Alignment-Free Nanometric Optical Metrology Enabled by Structured Light

Yu Wang[1], Benquan Wang[1], Eng Aik Chan[2], Songze Li[1], Zhenyu Guo[1], Xi Xie[3], Masako Kishida[4], Che-Chin Chen[5], Yijie Shen[1†], Jun-Yu Ou[6*]

*1. School of Physical and Mathematical Sciences, Nanyang Technological University, Singapore 637371, Singapore*

*2. Centre for Disruptive Photonic Technologies, Nanyang Technological University, Singapore 637371, Singapore*

*3. School of Physics, Chengdu University of Technology, Chengdu 610059, China*

*4. Faculty of Engineering, Information and Systems, University of Tsukuba, Tsukuba, Ibaraki 305-8573 Japan*

*5. National Applied Research Laboratories, Taiwan Instrument Research Institute, Hsinchu 30076, Taiwan*

*6. School of Physics and Astronomy, University of Southampton, SO17 1BJ, UK*

Corresponding authors Email: yijie.shen@ntu.edu.sg and bruce.ou@soton.ac.uk

## Abstract

Advances in the semiconductor industry are driven by the development of increasingly compact devices featuring intricate etched geometries, the characterization of which essentially requires ultraprecise, label-free, and real-time metrology. However, non-destructive and alignment-free optical metrology of sub-wavelength structures with nanometric resolution remains a major challenge. Here, we demonstrate a novel single-shot, label-free, and alignment-free optical metrology approach for determining the 1D position of sub-wavelength nanostructures, achieving λ/110 (7.2 nm) precision. The high precision benefits from utilizing structured illuminations of Laguerre–Gaussian (LG) or Hermite–Gaussian (HG) beams, and the AI analyzing method can retrieve the information when such structured light interacts with sub-wavelength objects. Furthermore, our approach leverages spatially distributed phase jumps in HG and LG beams interacting with the nanostructures, providing an alignment-robust solution to the challenges in optical metrology. Such an alignment-free, non-destructive, and high-precision metrology technique enables real-time machine vision, semiconductor inspection, and advanced manufacturing.

## Introduction

“If you can’t measure it, you can’t improve it.” stated by Lord Kelvin, emphasizing the significance of metrology [1]. Among them, the use of light-matter interactions to characterize properties of measurands is known as optical metrology [2-5]. Optical metrology has played a crucial role in advancing science and technology, particularly for high-precision measurements [6]. As early as the 19$^{th}$ Century, Michelson used interferometry to measure the wavelength of cadmium light, which was later used to define the meter before it was redefined in terms of the speed of light [7]. Today, optical interferometers continue to play a vital role in modern scientific research, enabling measurements in semiconductor inspection, astronomical observations, gravitational wave detection, etc. [8-10] In optical metrology, achieving both high precision and non-destructiveness has always been a key objective, particularly for observing the microscopic world [11]. It is well known that, as revealed by Abbe, features smaller than approximately half the wavelength of light cannot be resolved in conventional intensity-based microscopy [12]. Despite this theoretical limit, the modern fluorescent techniques offer the possibility to break such a boundary, such as stimulated emission depletion, photoactivated localization, and stochastic optical reconstruction microscopy, pushing the resolution to tens of nanometres [13-15]. However, the need for fluorescent labelling limits their applicability for non-invasive and real-time nanometrology. Rather than conventional illumination, structured light offers a solution to achieve non-destructive optical metrology [16, 17]. For instance, fringe-pattern light generated by fringe projection profilometry is demonstrated to achieve high-speed 3D dynamical shape measurement [16]. In phase measuring reflectometry, specular free-form surfaces can be rapidly and accurately measured by analyzing distortions in the reflected patterns [17].

Beyond this, even higher resolution, finer measurands, and broader application scenarios have been enabled by advances in computational algorithms, particularly machine learning [18-22]. Recently, neural networks have been able to measure the Brownian motion of a nanowire with negligible picometre error by learning diffraction fields [18]. Deep learning analysis of images of a virus-like nanoparticle illuminated by structured superoscillatory light enables high-speed super-resolution 3D localization [19]. Cross-domain learning of the sinusoidal patterns from surfaces enables accurate and adaptive 3D reconstruction of dynamic events [20]. Among them, the key aspect of machine learning in optical metrology, as a data-driven approach, is to establish the mapping between actual parameters of the measurand and the corresponding measured data. A more distinct correlation between measurable data and object parameters enhances the ability of the model to make more accurate predictions. In this context, although theoretical and experimental studies have demonstrated that structured light with phase singularities provides high accuracy for optical metrology [18, 23, 24], the necessity for precise alignment between the phase singularity and the measured objects inevitably introduces random statistical errors in practice [24]. Compared to the alignment of a phase singularity in optical metrology, spatially distributed phase jumps offer greater robustness for precision measurements. For example, structured light beams [25-27] such as Laguerre–Gaussian (LG) and Hermite–Gaussian (HG) beams with symmetric phase jumps can be effectively utilized as illumination sources in precision optical metrology, which not only enriches the measuring light sources but also reduces the difficulty of alignment.

Here, we demonstrate a single-shot, label-free, and alignment-free optical metrology framework for quantifying the one-dimensional position of sub-wavelength nanostructures. When sub-wavelength objects traverse LG or HG beams containing spatially distributed phase jumps, the resulting phase discontinuities generate displacement-sensitive intensity and phase signatures that can be retrieved using an artificial-intelligence-assisted inference model. In contrast to superoscillatory metrology schemes that rely on localized phase singularity, the use of distributed phase jumps in LG and HG beams relaxes alignment constraints while preserving high measurement precision. Experimentally, position measurement precisions of 7.2 nm (λ/110) and 7.5 nm (λ/106) are achieved under HG and LG illumination, respectively. This label-free high-precision metrology approach enables position measurements for both positive and negative nanostructures, providing a route towards applications in semiconductor inspection, nanophotonic device characterization, and real-time manufacturing monitoring.

## Principle and experimental setup

Recent theoretical and experimental investigations have shown that scattering patterns are highly sensitive to obstacles positioned at regions of giant phase gradients in structured light. Such findings imply that fine optical features interact more strongly with nanoscale objects when located in these high gradient regions, which significantly enhances the resolution and precision of optical microscopy and metrology. [18, 19, 28-30] A single slit is illuminated by the incident light $E_0(x,y) = A_0(x,y)\exp\big(i\varphi_0(x,y)\big)$, and a reflected far-field distribution $\big(E(u,v)\big)$ is given by the Fourier transform of the field across the aperture (the single slit):

$$E(u,v) \propto \mathcal{F}\{g(x,y)A_0(x,y)\exp\big(i\varphi_0(x,y)\big)\} \tag{1}$$

where $(x,y)$ and $(u,v)$ denote coordinates in the aperture (slit) plane and the far-field plane, respectively. $A_0(x,y)$ and $\varphi_0(x,y)$ are the amplitude and phase distributions of the incident light, respectively. The aperture function of the single slit $g(x,y)$ is the rectangular window function centred on the slit. A lateral position offset $\Delta x$ of the slit modifies the aperture function to $g(x+\Delta x,y)$, leading to the intensity of the reflected field recorded by the camera is:

$$I_\Delta(u,v) = |E(u,v)|^2 \propto |\mathcal{F}\{g(x+\Delta x,y)A_0(x,y)\exp\big(i\varphi_0(x,y)\big)\}|^2 \tag{2}$$

The intensity of the reflected field $I_\Delta(u,v)$ is not only determined by the local values of $A_0(x,y)$ and $\varphi_0(x,y)$ of the incident field but crucially by their spatial structure — that is, how they vary with position (i.e., their gradients). These steep gradients introduce high spatial-frequency components into the reflected field via the Fourier transform, leading to significant changes in the intensity pattern even for tiny perturbation (i.e., small displacement of the slit). Physically, this occurs because the local optical field acts like a sensitive probe: a small shift in the slit samples a completely different portion of the gradient field, altering the reflected outcomes. Therefore, under such structured illumination, the system exhibits extreme sensitivity to displacements of the slit and AI-enabled methodology extracts sensitive information from the reflected fields, facilitating ultra-precise optical metrology. Compared to phase singularities, phase jumps (infinite phase gradients) are theoretically more sensitive to the moving of nanometric objects and their accompanying spatial distributed giant intensity variations are more readily detected without the alignment to reduce measuring errors in optical metrology. Structured light beams such as Laguerre–Gaussian (LG) beams and Hermite–Gaussian (HG) beams, exhibiting doughnut-shaped and rectangularly symmetric phase jumps respectively, serve as effective illumination sources in our optical metrology approach. We utilize the structured LG or HG beam to illuminate the single slit, and the reflected field is collected by the camera, where the slit displaces along the x-axis ($\Delta x$). To enable such an optical metrology system to measure unknown slit position offsets ($\Delta x$) via machine learning, we first collect a training dataset comprising single-shot reflected images of the slit at 160 known positions and feed it into a neural network for training, and subsequently the trained network acquires the ability to accurately predict the slit's unknown position offsets, as illustrated in Fig. 1. Because the position information is encoded across multiple phase-transition regions rather than a single critical point (i.e., a phase singularity), the measurement is inherently alignment-free. Moreover, the phase topology of LG and HG beams remains largely preserved along propagation (apart from beam scaling and the Gouy phase shift), allowing the phase-jump distribution to persist across a range of axial planes, as shown in two green dashed planes in Fig. 1, without requiring a specific wavefront alignment.

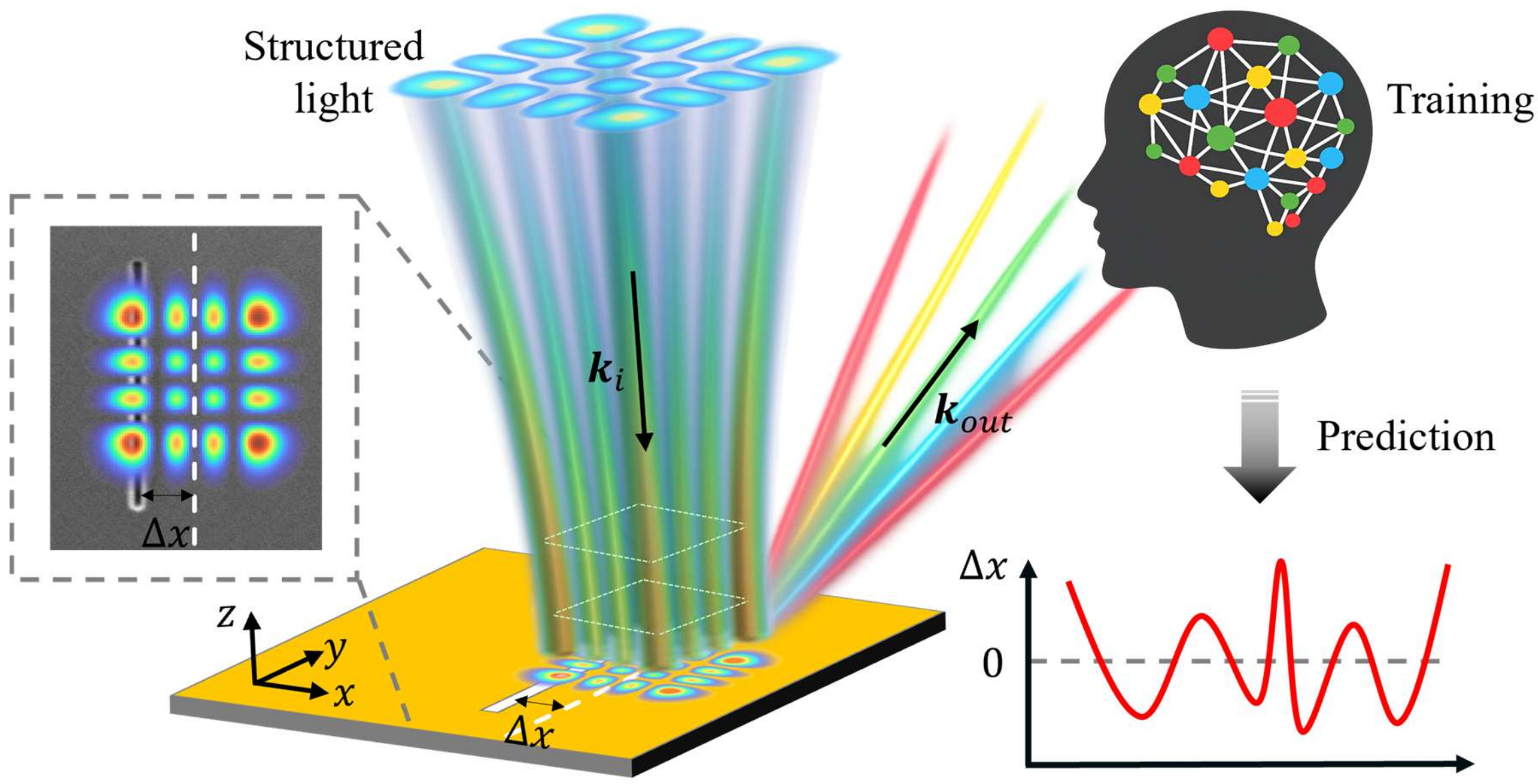

**Fig. 1 Conceptual schematic of nanometric optical displacement metrology enabled by structured light and machine learning. The incident structured light and the reflected field propagate along $K_i$ and $K_{out}$, respectively.**

Structured light can be produced by diffractive optical elements, metasurfaces, digital micromirror devices (DMDs), and spatial light modulators (SLMs) [26, 31-34]. We employ the SLM approach, enabling dynamic and reconfigurable beam shaping and offering high flexibility and precision in generating complex light fields, which is particularly advantageous for scanning and measuring objects in optical microscopy systems [34]. In the experimental setup, as shown in Fig. 2 (a), a Gaussian mode generated by a 795-nm laser is adjusted to linear polarization by sequentially passing through a linear polarizer (LP) and a half-wave plate (HWP). The beam is then expanded by two lenses (L1 and L2 with the focal lengths of 20 mm and 100 mm) with a magnification of 5, resulting in a near-plane wave. After passing through a beam splitter (BS), this near-plane wave illuminates a reflective phase-only SLM (Holoeye PLUTO), loaded with a hologram phase mask generated via the computer-generated holography (CGH) with blazed grating [35] to generate the tailored structured light. The modulated light field from the SLM is subsequently focused by a lens (L3) with an aperture (AP) positioned at the Fourier plane to select the first-order spatial pattern, forming the target structured light. An objective lens (100× and NA = 0.9) focuses the target field (LG or HG) onto the object (i.e., a single slit, which is placed on a piezo stage), and the reflected image of the object is collected by the camera through a tube lens (L5). In semiconductor inspection, positive slits (raised features) and negative slits (etched trenches) represent common elements in semiconductor alignment features, such as developed photoresist and etched alignment features. We fabricate two sub-wavelength single slits with length of 3 μm and width of 120 nm (8/53 λ), where one is a negative structure and another is a positive one, and their SEM images are displayed in Fig. 2 (b) and (c). A layer of chromium with thickness of 80 nm is deposited on a transparent glass substrate, and then the negative/positive slit is patterned by focused ion beam milling of the chromium layer inside/outside the rectangular area.

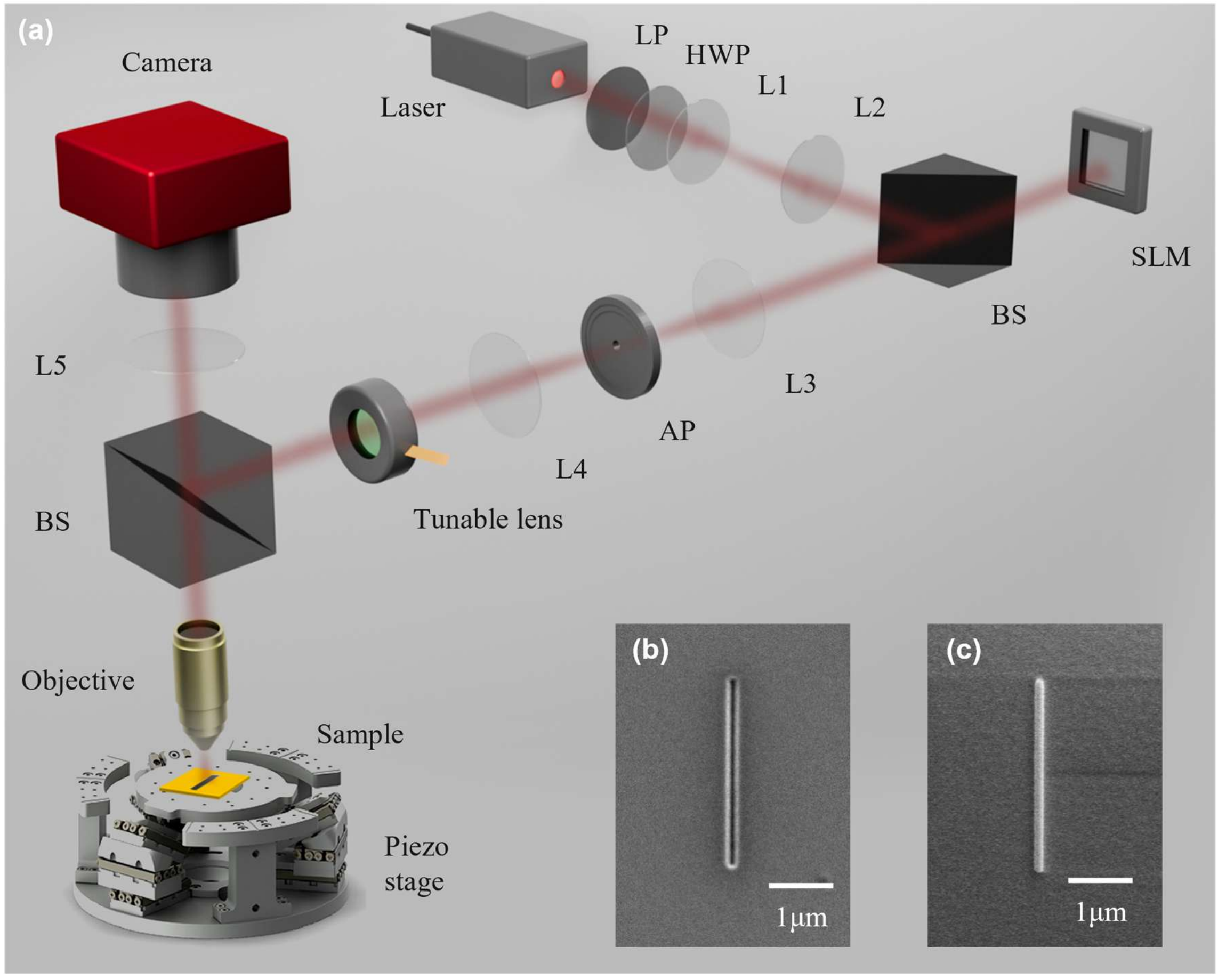

**Fig. 2 (a) Schematic diagram of the experimental setup. LP: linear polarizer, HWP: half-wave plate, L1-L5: lens, BS: beam splitter, SLM: spatial light modulator, AP: Aperture. (b and c) SEM images of two sub-wavelength single slits with negative and positive structures.**

A 34-layer residual network, following the architecture in reference [36], is employed to retrieve 1D position of the sub-wavelength single slit. This architecture is selected for its proven capability to efficiently learn complex mappings in imaging tasks such as object detection, classification, and segmentation [37-41]. The network begins with an initial convolutional layer utilizing a 7 × 7 kernel with 64 output channels, followed by a 3×3 max-pooling layer. It is structured into four stages containing 3, 4, 6, and 3 residual blocks, respectively. Within each stage, the residual blocks comprise multiple residual units, and each unit consists of 3 convolutional layers. A shortcut connection bypasses one or more layers within each unit, facilitating information preservation and alleviating the vanishing gradient problem. The rectified linear unit (ReLU) is used as the activation function throughout the network, except for the final operation within each block. Features from the last residual layer are aggregated via an average pooling layer, followed by a fully connected layer with 512 neurons and a final output neuron, which predicts 1D position of the single slit. The network is trained using the Adam stochastic optimization algorithm [42], with the objective of minimizing the mean absolute error loss function.

## Results and discussion

Here, one LG beam with radial and azimuthal indices of 3 and 0 ($LG_{3,0}$) and one HG beam with x- and y-axis nodes of 3 and 3 ($HG_{3,3}$) are constructed as structured illumination sources. To experimentally investigate the one-dimensional positional metrology of a sub-wavelength single slit, both negative and positive slits are illuminated by $LG_{3,0}$ or $HG_{3,3}$ beams. The reflected intensity field is recorded by a camera positioned at a distance of 2λ away from the object plane. The slit is displaced along the x-axis between -500 nm and 500 nm relative to the center of the illuminating spot with a spacing of 5 nm, and the ground-truth position is recorded by a piezoelectric stage. The neural network is trained using 160 reflection images collected from the slit at 160 different one-dimensional positions. Figures 3 (a1) and (b1) display representative reflection images of the negative and positive slits with illumination of $LG_{3,0}$ light where the slit is in the center of the illuminating light spot and Figs. 3 (c1) and (d1) show that of illumination of $HG_{3,3}$. After training, the neural network has the ability to retrieve the 1D position of the single slit directly from a previously unseen reflection image. The retrieval results are summarized in Figs. 3 (a2-d2), where the retrieved positions (black dots) are plotted against the true positions (blue line). Each data point corresponds to the mean value of 30 repeated predictions from the neural network to improve statistical stability and suppress the influence of outliers. The retrieval error is further quantified by $\Delta X = X_{R,i} - X_{\mathrm{T},i}$, as shown in Figs. 3 (a3-d3). The measurement precision is evaluated by the root-mean-square statistical error $\sigma = \sqrt{\frac{\sum_{i=1}^{n}\left(X_{R,i}-X_{\mathrm{T},i}\right)^2}{n}}$, where $X_{T,i}$ and $X_{R,i}$ denote the true position and the retrieved position, respectively, and $n$=21 refers to the number of measurements in the test dataset. Under $LG_{3,0}$ illumination in such optical metrology, the measurement precision (σ) reaches 12.2 nm (λ/65) for the negative slit and 7.5 nm (λ/106) for the positive slit, as shown in Figs. 3 (a2) and (b2). In addition, $HG_{3,3}$ illumination is also employed for the same metrology task, and the achieved measurement precision (σ) is 31.4 nm (λ/25) for the negative slit and 7.2 nm (λ/110) for the positive one in Figs. 3 (c2) and (d2), respectively.

For this reflective optical metrology system, the positive structure exhibits higher measurement precision under both $LG_{3,0}$ and $HG_{3,3}$ illuminations. Moreover, $HG_{3,3}$ illumination provides superior performance for the positive slit because its Cartesian-symmetry phase and intensity distributions generate stronger direction-dependent spatial features during one-dimensional displacement, which can be more effectively decoded by the neural network. These nanometer-scale measurement precisions originate from the intrinsic phase-jump structures of $LG_{3,0}$ and $HG_{3,3}$ beams, which greatly enhance the sensitivity of the reflected optical field to nanoscale perturbations, thereby enabling accurate optical metrology of subwavelength structures.

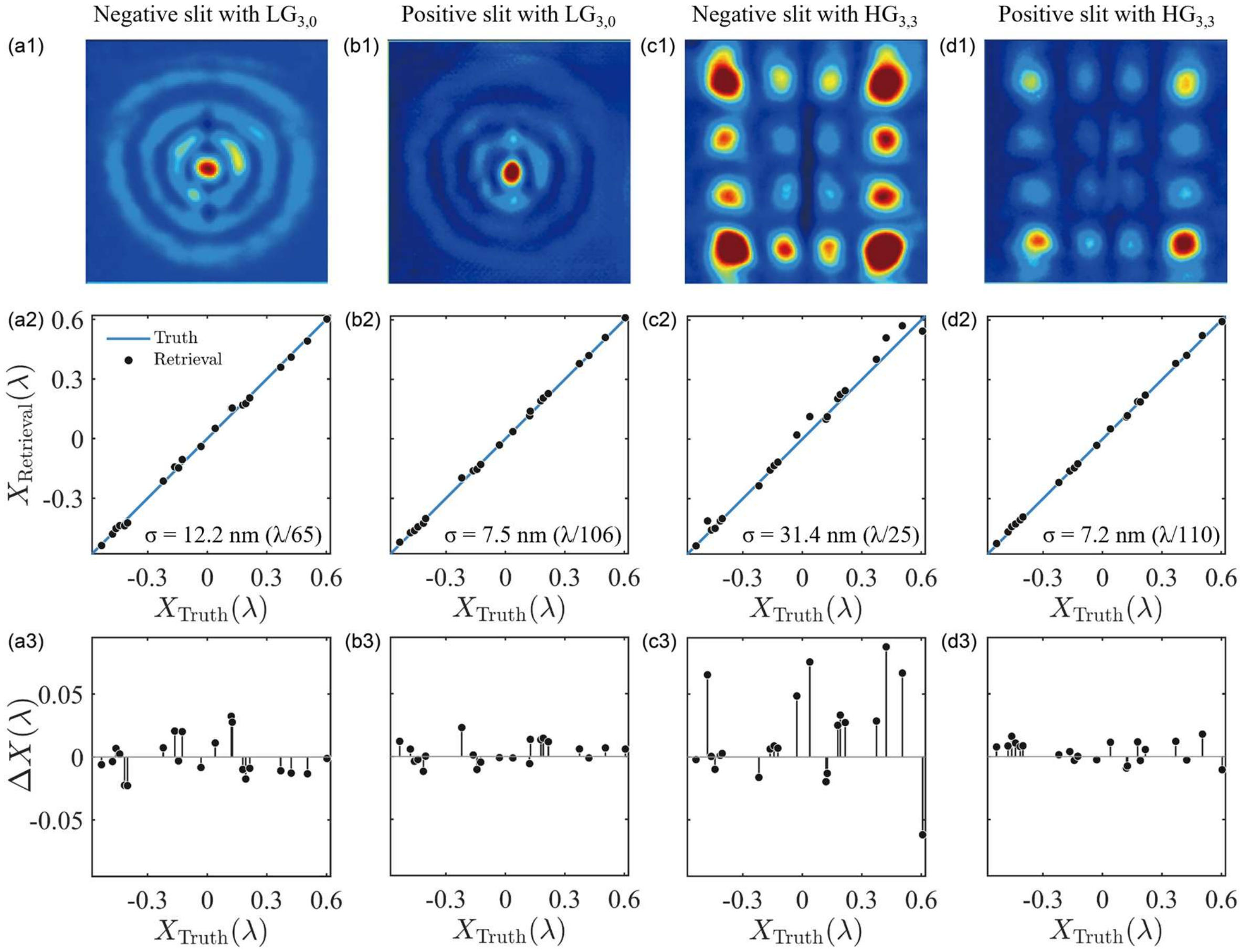

**Fig. 3 Experimental 1D positional metrology of a sub-wavelength single slit based on structured light illumination and machine learning. (a1, b1) Representative reflection images of the negative and positive single slits illuminated by the $LG_{3,0}$ beam. (c1, d1) Representative reflection images of the negative and positive single slits illuminated by the $HG_{3,3}$ beam, respectively. (a2–d2) Retrieved 1D positions vs. the true positions under $LG_{3,0}$ and $HG_{3,3}$ illuminations, where black dots represent the positions retrieved by the neural network while the blue solid lines denote the**

**truth. The corresponding measurement precision σ is indicated in each panel. (a3–d3) The error between the measured positions and the true positions in the test dataset.**

Figure 4 displays the target and experimentally retrieved phase distributions and local wavevector ($k$) profiles of illuminating $LG_{3,0}$ and $HG_{3,3}$ light fields. In the experiment, the optical phase is retrieved by an interferometric scheme implemented within a Stokes measurement framework [43]. A right-handed circularly polarized beam encoding the target phase ($LG_{3,0}$ or $HG_{3,3}$) serves as the object, while a left-handed circularly polarized plane wave acts as the reference. The two beams are combined coaxially in a Sagnac interferometer and projected onto four linear polarization bases to record interferometric intensities $I_H$, $I_V$, $I_A$, and $I_D$, together with the intensity of the object beam $I_R$ and that of the reference beam $I_L$. Here, $I_H$, $I_V$, $I_D$, and $I_A$ denote the intensities measured after projection onto the horizontal, vertical, diagonal (+45), and anti-diagonal (−45) linear polarization bases, respectively, while $I_R$ and $I_L$ correspond to the intensities of the right-handed circularly polarized object beam and the left-handed circularly polarized reference beam. The object phase is then directly reconstructed using a standard four-step relation, $\varphi = arctan\frac{I_D - I_A}{I_H - I_V}, \varphi \in (-\pi, \pi]$. The experimentally reconstructed phase map of $LG_{3,0}$ beam in Fig. 4 (a2) exhibits a series of concentric annular phase domains, where adjacent rings are separated by abrupt $\pi$-level phase discontinuities, giving rise to radially distributed phase jumps with clear rotational symmetry about the beam axis, which clearly reproduces the annular phase-jump structure predicted in the simulation, with only minor distortions arising from experimental noise and residual aberrations. The experimental phase distribution agrees well with the simulation in Fig. 4 (a1), with only minor deviations arising from noise and residual aberrations. In order to quantify the local phase variation of the illuminating beam, we compute the spatial gradient of the phase field (i.e., local k). Figure 4 (a3) shows the resulting 2D map of |local k|, in which concentric ridges appear at the locations of phase discontinuities. These ridges correspond to regions where the phase varies extremely rapidly, reflecting the sharp phase jumps inherent to the $LG_{3,0}$ mode. To visualize this more clearly, a horizontal cross-section along the red dashed line is plotted in Fig. 4 (a4). The cross-section reveals a series of narrow and high peaks reaching values on the order of more than 90 $k_0$. Using the same four-step phase-shifting reconstruction procedure, we also retrieve the phase of the $HG_{3,3}$ illumination. The experimentally retrieved phase distribution in Fig. 4 (b2) exhibits a checkerboard-like structure composed of orthogonally arranged phase domains. Across each horizontal and vertical boundary, the phase undergoes abrupt $\pi$-level discontinuities, forming a grid of sharply defined phase jumps, which matches well with the target on in Fig. 4 (b1). The cross-section of the local wavevector shown in Fig. 4 (b4) exhibits a series of narrow and high peaks approaching 80 $k_0$. These pronounced peaks in both $LG_{3,0}$ and $HG_{3,3}$ indicate that the network of phase jumps is extremely sensitive to 1D/2D position shifts. As the illumination scans across the nano-object, the nano-object is swept by these phase-jump boundaries, producing abrupt, large variations in the far-field signal. This strong phase–sensitivity makes the configuration highly suitable for ultra-precise optical metrology.

The demonstrated distributions of phase jumps and local wave vectors also suggest a practical criterion for mode-order selection. Higher-order LG or HG modes can provide more phase-jump boundaries, potentially improving displacement sensitivity [44]. Nevertheless, very high-order modes may reduce experimental robustness because of imperfect mode generation, optical aberrations, lower local intensity, and more complex image features for neural-network inference. Thus, $LG_{3,0}$ and $HG_{3,3}$ are used here as representative intermediate-order modes to demonstrate the feasibility of the proposed approach, while the optimal order for a specific metrology task should be selected by numerical simulation, experimental calibration, and machine-learning validation.

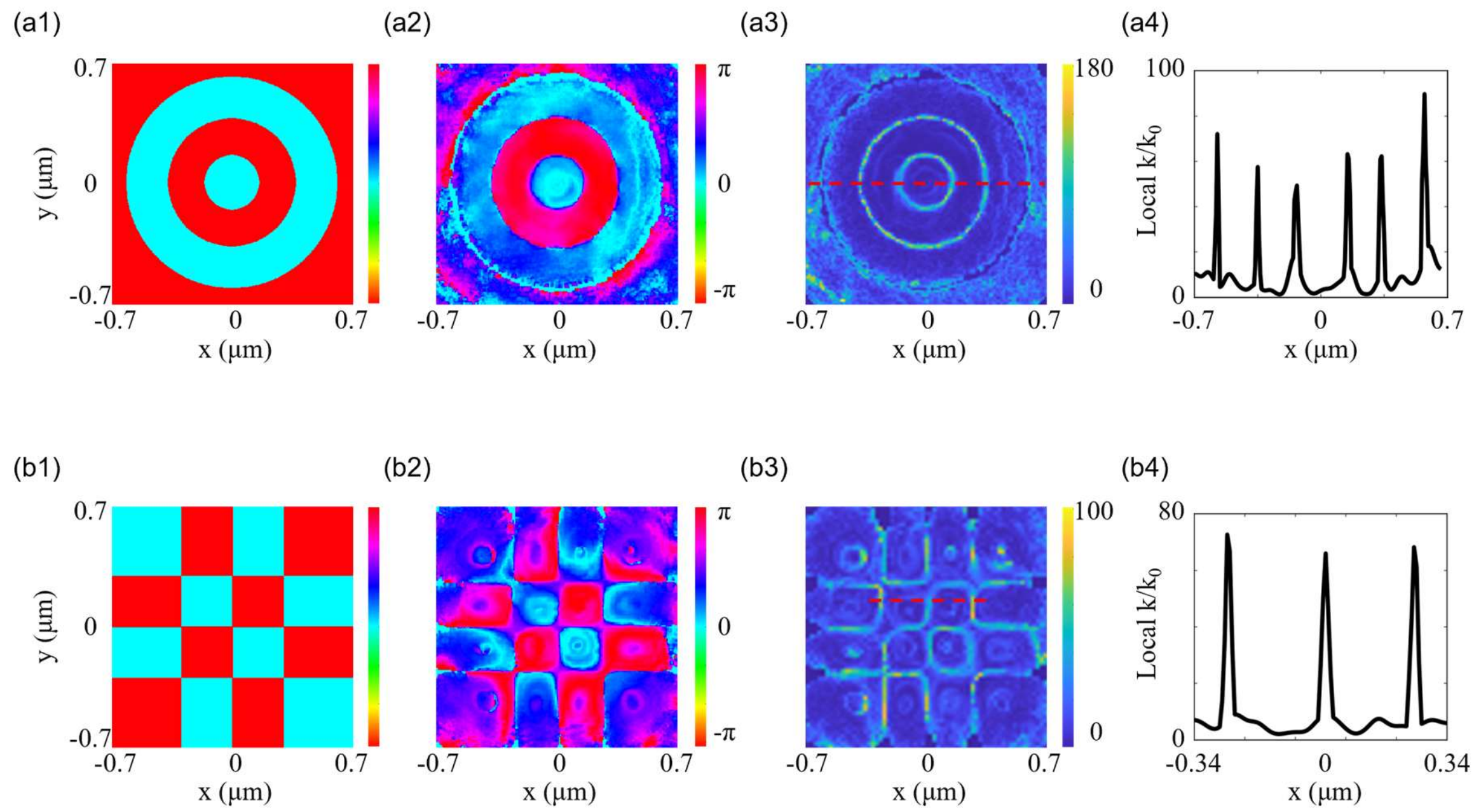

**Fig. 4 The simulated (a1 and b1) and experimentally obtained (a2-a4 and b2-b4) structured illumination $LG_{3,0}$ and $HG_{3,3}$ light. (a1 and b1) The simulated phase profiles of $LG_{3,0}$ and $HG_{3,3}$. (a2 and a3) The experimental phase distribution and local wave vector (k) of $LG_{3,0}$. (a4) The cross-sectional local k along the red dashed line in (a3). (b2and b3) The experimental phase distribution and local k of $HG_{3,3}$. (b4) The cross-sectional local k along the red dashed line in (b3).**

## Conclusion

In summary, accurate 1D positional nanometrology with deep sub-wavelength precision has been demonstrated experimentally, which uses a neural network to retrieve the 1D displacement of a sub-wavelength object from its image recorded with LG or HG illumination. This technique can locate nano-objects within milliseconds, i.e., within the time it takes to record a single-shot image with a camera. The metrology precision surpassing the Abbe–Rayleigh diffraction limit in conventional microscopy dozens of times has been demonstrated here on a system allowing for the collection of physical training data for the neural network. Compared with an object with a negative structure, such a reflective metrology system can measure the positive-structure object more accurately. More importantly, the technique expands the possibility of using other types of structured light with phase jumps as illuminations in optical metrology, machine vision, and semiconductor inspection.

**Research funding:** This work was supported by the UK's Engineering and Physical Sciences Research Council (project numbers: EP/X041166/1, EP/T02643X/1, and UKRI257). J. Y. Ou acknowledges support from the Department of Science, Innovation and Technology's International Science Partnerships Fund (ISPF) via the Royal Academy of Engineering under the Frontiers programme. Singapore Ministry of Education (MOE) AcRF Tier 1 (RG157/23 & RT11/23), Singapore Agency for Science, Technology and Research (A*STAR) (M24N7c0080 & 256I9013), and Nanyang Assistant Professorship Start Up grant.

**Author contribution:**

Y. Wang and J. Y. Ou conceived, planned, and carried out the experiments. B. Wang and E. A. Chan constructed the neural network. J. Y. Ou prepared the sample. S. Li, X. Xie, and Z. Guo contributed to structured light field generation and analysis. M. Kishida and C.C. Chen provided suggestions on the neural network and instrumentation. Y. Shen and J. Y. Ou supervised the work and contributed to the interpretation of the results. All authors provided critical feedback and helped shape the research, analysis, and manuscript. The authors are grateful to Y. Shen and Nikolay I. Zheludev for providing the experimental equipment.

**Conflict of interest:**

The authors declare no conflict of interest.

**Data availability statement:**

The data that support the findings of this study are openly available in the University of Southampton ePrints research repository at https://doi.org/10.5258/SOTON/xxxxxx.